\documentclass[11pt,oneside,english]{amsart}
\usepackage{amssymb}

\makeatletter
\usepackage{amsmath,amsthm}
\usepackage{amsfonts}
  \theoremstyle{plain}
  \newtheorem*{thm*}{Theorem}
  \theoremstyle{remark}
  \newtheorem*{acknowledgement*}{Acknowledgement}

\makeatother

\begin{document}

\title{An extension of Wick's theorem}

\author{C. Vignat and S. Bhatnagar}

\address{Institut Gaspard Monge, Universit\'{e} de Marne la Vall\'{e}e, France and
Department of Computer Science and Automation, Indian Institute of
Sciences, Bangalore, India}

\email{vignat@univ-mlv.fr, shalabh@csa.iisc.ernet.in}

\begin{abstract}
We propose an extension of a result by Repetowicz et al. \cite{repetowicz}
about Wick's theorem and its applications: we first show that Wick's
theorem can be extended to the uniform distribution on the sphere
and then to the whole class of elliptical distributions. Then,
as a special case, we detail this result for distributions that are
Gaussian scale mixtures. Finally, we show that these results allow
to recover easily a theorem by Folland \cite{folland} about integration
of polynomials over the sphere.
\end{abstract}
\maketitle

\section{Wick's theorem}

In quantum field theory \cite{etingof}, the determination of the
partition function of some systems involves the computation of integrals
of the form \[
\int_{\mathbb{R}^{n}}P\left(X\right)\exp\left(-\frac{X^{t}\Sigma^{-1}X}{2}\right)dX,\]
where $P\left(X\right)$ is a monomial in the components of $X\in\mathbb{R}^{n}$,
that is, in a probabilistic approach, mixed moments of a Gaussian
random vector with covariance matrix $\Sigma.$ 

Wick's theorem provides a simple formula for these moments: let us
denote by $X$ a Gaussian vector in $\mathbb{R}^{n}$ with zero mean
and covariance matrix $\Sigma;$ a linear form ${\normalcolor \mathcal{L}}_{i}$
on $\mathbb{R}^{n}$ is identified with the vector of its coefficients
\[
\mathcal{L}_{i}\left(X\right)=l_{i}^{t}X\]
 A \textit{pairing} $\sigma$ on the set $I_{2m}=\left\{ 1,\dots,2m\right\} $
is a partition of $I_{2m}$ into $m$ disjoint pairs; the set $\Pi_{2m}$
of pairings of $I_{2m}$ is composed of $\frac{2m!}{2^{m}m!}$ elements.
For example, in the case $m=2,$ there are $3$ different pairings
on $I_{4}=\left\{ 1,2,3,4\right\} ,$ namely $\sigma_{1}=\left\{\left(1,2\right),\left(3,4\right)\right\},$
$\sigma_{2}=\left\{\left(1,3\right),\left(2,4\right)\right\}$ and $\sigma_{3}=\left\{\left(1,4\right),\left(2,3\right)\right\}$.
We denote by $I_{2m}/\sigma$ the set of indices $i$ such that $\sigma=\left\{\left(i,\sigma\left(i\right)\right)\right\}.$

Wick's theorem states as follows \cite{etingof}.

\begin{thm*}
\label{thm:Wick}Let $m$ be an integer then\[
E\prod_{i=1}^{2m}\mathcal{L}_{i}\left(X\right)=\sum_{\sigma\in\Pi_{2m}}\prod_{i\in I_{2m}/\sigma}l_{i}^{t}\Sigma l_{\sigma\left(i\right)}\]
 If $2m$ is replaced by $2m+1$, then the expectation above equals
zero. 
\end{thm*}
As an example, with $m=2$ and $\mathcal{L}_{i}\left(X\right)=X_{i},$
we obtain \[
E\left[X_{1}X_{2}X_{3}X_{4}\right]=E\left[X_{1}X_{2}\right]E\left[X_{3}X_{4}\right]+E\left[X_{1}X_{3}\right]E\left[X_{2}X_{4}\right]+E\left[X_{1}X_{4}\right]E\left[X_{2}X_{3}\right].\]

\section{Spherical and Elliptical distributions}

In this paper, we extend Wick's theorem to the class of spherical
random vectors: such a vector $X\in\mathbb{R}^{n}$ is characterized
\cite[Ch.2]{fang} by the fact that its characteristic function $\psi\left(Z\right)=E\exp\left(iZ^{t}X\right)$
is invariant by orthogonal transformation; in other words, \[
\psi\left(Z\right)=\phi\left(Z^{t}Z\right)\]
for some function $\phi:\mathbb{R}^{+}\rightarrow\mathbb{C}.$ In
this case, a stochastic representation of $X$ writes\[
X=rU\]
where $U$ is a random vector uniformly distributed on the unit sphere
$\mathcal{S}_{n}=\left\{ X\in\mathbb{R}^{n}; \Vert X\Vert=1\right\} $
and $r\in\mathbb{R}^{+}$ is a positive random variable independent
of $U.$ If moreover vector $X$ admits a density $f_{X},$ then it
is necessarily of the form\[
f_{X}\left(X\right)=g\left(X^{t}X\right)\]
 for some function $g:\mathbb{R}\rightarrow\mathbb{R}^{+}.$ 

An extension of the spherical property is the elliptical property
\cite[Ch.2]{fang}: a random vector $Y\in\mathbb{R}^{n}$ is elliptical
with $\left(n\times n\right)$ characteristic matrix $\Sigma$ if
\begin{equation}
Y=\Sigma^{1/2}X\label{eq:elliptic}\end{equation}
where $X$ is spherical. We assume without loss of generality that
all vectors have zero mean.

\section{Wick's theorem for the uniform distribution on the sphere}

Wick's theorem can be extended to the case of a vector uniformly distributed
on the sphere $\mathcal{S}_{n}$ as follows.

\begin{thm*}
If $U$ is uniformly distributed on the sphere $\mathcal{S}_{n}$
then\[
E\prod_{i=1}^{2m}\mathcal{L}_{i}\left(U\right)=\frac{\Gamma\left(\frac{n}{2}\right)}{2^{m}\Gamma\left(m+\frac{n}{2}\right)}\sum_{\sigma}\prod_{i\in I_{2m}/\sigma}l_{i}^{t}l_{\sigma\left(i\right)}\]

\end{thm*}
\begin{proof}
A stochastic representation for vector $U$ is\[
U=\frac{N}{\Vert N\Vert}\]
where $N$ is a Gaussian vector with unit covariance matrix. By the
polar factorization property, random variable $r=\Vert N\Vert$ is
independent of $U;$ we deduce from the linearity of $\mathcal{L}_{i}$
that \begin{eqnarray*}
E\prod_{i=1}^{2m}\mathcal{L}_{i}\left(N\right)=E\prod_{i=1}^{2m}r\mathcal{L}_{i}\left(U\right) & = & Er^{2m}E\prod_{i=1}^{2m}\mathcal{L}_{i}\left(U\right)\end{eqnarray*}
so that \[
E\prod_{i=1}^{2m}\mathcal{L}_{i}\left(U\right)=\frac{E\prod_{i=1}^{2m}\mathcal{L}_{i}\left(N\right)}{Er^{2m}}.\]
Since $r=\sqrt{N^{t}N}$ is chi-distributed with $n$ degrees of freedom,
elementary algebra yields\[
Er^{2m}=\frac{2^{m}\Gamma\left(m+\frac{n}{2}\right)}{\Gamma\left(\frac{n}{2}\right)}.\]

\end{proof}
From the linearity of forms $\mathcal{L}_{i}$ and by stochastic representation
(\ref{eq:elliptic}), we deduce a version of Wick's theorem for any
elliptical vector as follows.

\begin{thm*}
If $X\in\mathbb{R}^{n}$ is an elliptical vector with stochastic representation
\begin{equation}
X=a\Sigma^{1/2}U\label{eq:XaU}\end{equation}
and if random variable $a$ has moments up to order $2m,$ then\[
E\prod_{i=1}^{2m}\mathcal{L}_{i}\left(X\right)=\frac{Ea^{2m}}{Er^{2m}}\sum_{\sigma}\prod_{i\in I_{2m}/\sigma}l_{i}^{t}\Sigma l_{\sigma\left(i\right)}\]

\end{thm*}

\section{Wick's theorem for Gaussian scale mixtures}

A random vector $X$ is a Gaussian scale mixture if its probability distribution
writes\[
f_{X}\left(X\right)=\int_{0}^{+\infty}g\left(x;a\Sigma\right)dH\left(a\right)\]
 where $g\left(x;\Sigma\right)$ is the Gaussian distribution with $\left(n\times n\right)$ covariance matrix $\Sigma$ on $\mathbb{R}^{n}$ 
and $H$ is a cumulative distribution function. A stochastic
representation of such a vector is\begin{equation}
X=\frac{N}{\sqrt{a}}\label{eq:scalemixture}\end{equation}
 where $N$ is a Gaussian vector with covariance matrix $\Sigma$
and $a$ is a scalar random variable independent of $N$ with cumulative
distribution function $H.$ Cauchy distributions, alpha-stable and
Student-t distributions belong to the family of Gaussian scale mixture
distributions. 

If $X$ is a Gaussian scale mixture, then it is also elliptical; however,
the contrary is not true, as shown \cite{chu} by the Pearson II (or
Student-r) family of distributions\[
f_{X}\left(X\right)=\frac{\Gamma\left(\frac{2-q}{1-q}+\frac{n}{2}\right)}{\Gamma\left(\frac{2-q}{1-q}\right)\vert\pi pK\vert^{1/2}}\left(1-\frac{X^{t}K^{-1}X}{p}\right)_{+}^{\frac{1}{1-q}}\]
 with $p=n+2\frac{2-q}{1-q}>0$ and notation $\left(x\right)_{+}=\max\left(x,0\right).$ 

Using stochastic representation (\ref{eq:scalemixture}), Wick's theorem
extends to Gaussian scale mixtures as follows.

\begin{thm*}
If $X$ is a Gaussian scale mixture as in (\ref{eq:scalemixture})
with $Ea^{-m}<\infty,$ then \[
E\prod_{i=1}^{2m}\mathcal{L}_{i}\left(X\right)=E\left[a^{-m}\right]\sum_{\sigma\in\Pi_{m}}\prod_{i\in I_{2m}/\sigma}l_{i}^{t}\Sigma l_{\sigma\left(i\right)}\]
 
\end{thm*}
We note that Repetowitz et al. \cite{repetowicz} deal only with the
case of Student-t distributions for which random variable $a$ is
Gamma distributed; the proposed theorem extends thus their result
to the class of all Gaussian scale mixtures for which $Ea^{-m}<\infty$.

\section{Application: a result by Folland}

In this section, we show that the results above allow to recover easily
the following theorem by Folland \cite{folland} about integration
of a monomial over the sphere $\mathcal{S}_{n}.$

\begin{thm*}
If \[
P\left(X\right)=\prod_{i=1}^{n}X_{i}^{\alpha_{i}}\]
then\[
\int_{\mathcal{S}_{n}}P\left(X\right)d\sigma=\begin{cases}
0 & \text{if some }\alpha_{i}\text{ is odd}\\
\frac{2\Gamma\left(\beta_{1}\right)\dots\Gamma\left(\beta_{n}\right)}{\Gamma\left(\beta_{1}+\dots+\beta_{n}\right)} & \text{else}\end{cases}\]
where\[
\beta_{i}=\frac{\alpha_{i}+1}{2}\]
and $d\sigma$ is the surface measure on the sphere $\mathcal{S}_{n}.$
\end{thm*}
\begin{proof}
Defining $\alpha_{0}=0,$ we consider the function $T:\left[1,\sum_{i=1}^{n}\alpha_{i}\right]\rightarrow\left[1,n\right]$
such that \[
T\left(i\right)=j \,\,\,\, \text{if } \,\,\,\, \alpha_{1}+\dots+\alpha_{j-1}<i\le\alpha_{1}+\dots+\alpha_{j}.\]
We choose forms $\mathcal{L}_{i}$ such that $l_{i}=\delta_{T\left(i\right)}$ where $\delta_{k}$
is the k-th column vector of the $n\times n$ identity matrix. Since the
scalar product $l_{i}^{t}l_{\sigma\left(i\right)}$ equals zero when
$T\left(i\right)\ne T\left[\sigma\left(i\right)\right],$  we consider only the $\prod_{i=1}^{n}\frac{\alpha_{i}!}{2^{\frac{\alpha_{i}}{2}}\left(\frac{\alpha_{i}}{2}\right)!}$
pairings for which $T\left(i\right)=T\left[\sigma\left(i\right)\right]$
and $l_{i}^{t}l_{\sigma\left(i\right)}=\delta_{T\left(i\right)}^{t}\delta_{T\left(i\right)}=1.$
Thus, by Wick's theorem,\[
E\prod_{i=1}^{n}\mathcal{L}_{i}\left(U\right)=\frac{\Gamma\left(\frac{n}{2}\right)}{2^{m}\Gamma\left(m+\frac{n}{2}\right)}\prod_{i=1}^{n}\frac{\alpha_{i}!}{2^{\frac{\alpha_{i}}{2}}\left(\frac{\alpha_{i}}{2}\right)!}=\frac{\Gamma\left(\frac{n}{2}\right)}{2^{m}\Gamma\left(m+\frac{n}{2}\right)}\frac{2^{\sum_{i=1}^{n}\beta_{i}}}{\left(2\pi\right)^{n/2}}\prod_{i=1}^{n}\Gamma\left(\beta_{i}\right).\]
Since $d\sigma$ is the unnormalized surface measure and as the
surface of the sphere is $S\left(\mathcal{S}_{n}\right)=\frac{2\pi^{n/2}}{\Gamma\left(\frac{n}{2}\right)},$\[
\int_{\mathbb{R}^{n}}P\left(X\right)d\sigma=S\left(\mathcal{S}_{n}\right)\times E\prod_{i=1}^{n}\mathcal{L}_{i}\left(U\right)\]
which yields the result.
\end{proof}
\begin{acknowledgement*}
This work was performed during a visit by C. Vignat to S. Bhatnagar
to the Computer Science and Automation Department of the Indian Institute
of Sciences, Bangalore. C.V. thanks S. B. for his nice welcome.
\end{acknowledgement*}

\end{document}